\documentclass[11pt]{article}

\usepackage[margin=1in]{geometry}

\newcommand{\proptit}{Carroll Mechanisms:\\Opportunities, Challenges, and Agenda}

\usepackage{sgame}
\usepackage{tablefootnote}
\usepackage{amsthm}
\usepackage{amssymb}
\usepackage{amsfonts}
\usepackage{amsmath}
\usepackage{graphics}
\usepackage{graphicx}
\usepackage{epsfig}
\usepackage{epstopdf}
\usepackage{mathrsfs}
\usepackage{cite}
\usepackage[mathscr]{euscript}
\usepackage{float}
\usepackage{wrapfig}
\usepackage[shortlabels]{enumitem}
\usepackage[dvipsnames]{xcolor}
\usepackage[font=small,labelfont=bf]{caption}
\usepackage{subcaption}
\usepackage[bottom]{footmisc}
\newlength{\mdframemarg}
\usepackage[	linewidth=.25mm,
					skipabove=0mm,
					innerleftmargin=\mdframemarg,
					innerrightmargin=\mdframemarg,
					innertopmargin=.5\mdframemarg,
					innerbottommargin=\mdframemarg]{mdframed}
\usepackage{hyperref}
\hypersetup{
  colorlinks=true,
  citecolor=Green,
  urlcolor=blue!70!green
}
\usepackage{tikz}
\usetikzlibrary{positioning}
\usetikzlibrary{calc}

\newcommand{\notetoself}[1]{}

\newcommand{\q}{\mathbf{q}}

\usepackage{setspace}

\newcommand{\mycite}[1]{~\cite{1}}

\newtheorem{theorem}{Theorem}

\newtheorem{proposition}[theorem]{Proposition}

\newcounter{oqcounter}

\newcommand{\addperiod}[1]{#1}

\usepackage{titlesec}
\titleformat{\section}{\normalfont\Large\bfseries}{\thesection.}{0.5em}{\addperiod}
\titleformat{\subsection}{\normalfont\large\bfseries}{\thesubsection}{.5em}{}
\titleformat{\subsubsection}{\normalfont\normalsize\bfseries}{\thesubsubsection}{0.5em}{}

\titlespacing*{\section}{0mm}{4mm}{1.5mm}
\titlespacing*{\subsection}{0mm}{3mm}{1.5mm}
\titlespacing*{\subsubsection}{0mm}{2.5mm}{1mm}

\begin{document}

\begin{center}
\Large\proptit\footnote{This research was funded by the Scroll DAO under proposal FY25-RnDAO.}

\quad\\

\normalsize Philip N. Brown\footnote{Dr. Brown (\href{mailto:philipnbrown@gmail.com}{\tt philipnbrown@gmail.com}) participated in this research as a paid consultant with the Network Goods Institute; he is also an Associate Professor of Computer Science at the University of Colorado Colorado Springs, USA.} and Connor McCormick\footnote{Connor McCormick (\href{mailto:cnnr.mccrmck@gmail.com}{\tt cnnr.mccrmck@gmail.com}) is with the Network Goods Institute (\url{https://www.networkgoods.institute/}).}%

\end{center}

\setcounter{tocdepth}{3} 
\tableofcontents

\pagestyle{plain}

\section{Introduction}
The purpose of Carroll Mechanisms is to facilitate autonomous group sensemaking and reasoned decisionmaking by incentivizing participants to be transparent about their reasoning process.
We envision Carroll Mechanisms to be built on top of a networked combinatorial LMSR foundation and thus to inherit the desriable properties of market scoring rules and automated market-makers.
While we have made great strides during Fall 2025 in building out this foundation (see summary in Section~\ref{sec:foundation}), several significant questions remain and several major new questions have arisen as a result of this work.
The purpose of this document is to frame these questions clearly and propose a research plan to address the questions.

While this document contains much detail and covers many issues, I argue that the most important pieces of a future research plan are:
\begin{enumerate}\label{list: 3 key pieces}
    \item A clear and concrete statement of system objectives: What problem do we want this work to solve, and why don't incumbent systems solve this problem already?
    \item A clear and concrete meta-architecture: for instance, does the system take action (disburse funding to selected stakeholders) or is it purely advisory? 
    Does the system operate under a deadline model, under a quiescence model, or something else?
    Is the system literally an extension of Futarchy, or does it simply operate in a similar vein?
    \item A deeper, more extensive literature review to understand the ways that others have approached this and related problems.
\end{enumerate}

This document should serve to seed the creation of those three pieces by clearly documenting the progress and the shortcomings of the work we have accomplished so far.
The Fall 2025 project focused entirely on a novel networked combinatorial LMSR-based approach (Section~\ref{sec:foundation}) with the added affordances of Restake, Doubt, and Mindchange (Section~\ref{sec: EL}).
The combinatorial LMSR backbone enables a wide variety of functionality and initially appeared to be an extremely promising avenue for development.
In addition, the prior work which led to the project in the first place had a clear picture of the Restake, Doubt, and Mindchange mechanics which significantly influenced us to proceed in that direction. 
Thus, at the beginning of the project, we believed that an implementation/development-heavy approach was warranted due to our prior understanding of preliminary work.
However, we were unable to find an implementation of these affordances which appropriately integrated into the LMSR framework in a way that accomplished our goals without also introducing damaging attack vectors or failure modes.

In addition, our technical and conceptual foundation has some unresolved issues:
\begin{enumerate}
    \item There are conceptual concerns about the need for the Doubt/Mindchange mechanic: see Section~\ref{ssec: doubt and mindchange mechanism}.
    \item The technical combinatorial LMSR framework does not currently have a means to resolve the \textit{relevance} between two propositions and we have no particularly promising leads except \textit{possibly} self-resolving prediction markets (see Section~\ref{sssec: self-resolving}).
    \item Similarly, the LMSR framework does not currently render decisions in a principled way (i.e., all nodes and edges are effectively modeled as resolvable prediction markets). 
\end{enumerate}

Due to these shortcomings and the lack of satisfying answers fo the main questions posed on Page~\pageref{list: 3 key pieces}, we feel comfortable recommending that this is an appropriate moment in the process to ``zoom out'' and address the three key pieces listed at the top of Page~\pageref{list: 3 key pieces}.
For an extended discussion of our current understanding of that list, see Section~\ref{ssec: the top list}.
The overall document structure places a brief literature review in Section~\ref{sec: related work}, the technical foundation in Section~\ref{sec:foundation}, a close look at the goals and issues of the Epistemic Leverage problem in Section~\ref{sec: EL}, and a comprehensive research plan in Section~\ref{sec: research agenda}.

\section{Related Work}
\label{sec: related work}

What follows is a brief selection of some relevant papers; this is not meant to be exhaustive but rather can be used as a starting point for the much longer task of conducting an extensive literature review.

First, here is a very nice (but old) survey on prediction markets from 2010~\cite{Chen2010}.
It is a fine place to begin learning about these systems.
Another good place to begin is with some of the earlier Hanson work, such as~\cite{Hanson2003}.
Central to this document is the notion of Combinatorial LMSR; my favorite exposition is~\cite{Chen2008a}, but there is other information in~\cite{Hanson2003,Hanson2007,Pennock2011} and very recent work as well~\cite{Hossain2025}.

Decision markets are an important setting for us; however, using a prediction-market-like mechanism to render real-world decisions generally appears to be a hard problem.
For example, Othman and Sandholm~\cite{Othman2010} show a universal ``manipulation'' modality in certain types of decision markets whereby a neutral fully-informed trader may gain personally by causing a suboptimal outcome to be selected.
Yiling Chen has a nice follow-on paper~\cite{Chen2011}, and more on requiring mixed strategies in decision-making in their journal paper~\cite{Chen2014a}.

Outside of decision markets but still relevant are settings in which the market predictions have some effect on the points being predicted; for example if voters respond to predictions about the outcomes of an election.
There is a reasonable quantity of work on this type of problem; for example, here is a setting where good equilibria only occur when some decision-makers (relative to the point under discussion) are unlikely to participate in the market~\cite{Chakraborty2016}.
Here is another where they discuss ``performative'' prediction; i.e., predictions which influence the outcome of events~\cite{Oesterheld2023}.

Finally, this very recent paper is of interested to our ``relevance elicitation'' problem; it proposes the first incentive compatible self-resolving prediction markets for unverifiable outcomes~\cite{Srinivasan2025}.
The proposed mechanism's incentive compatibility results depend on some assumptions which may fail in our setting, but there is a chance that this mechanism may provide some opportunity for us to perform relevance elicitation, or even possibly to render governance decisions.
We expand upon this in Section~\ref{sssec: self-resolving}, where we conclude (among other things) that the common prior assumption is necessary for self-resolving prediction markets to be truthful.
There may be some possibility that they remain useful for our purposes despite lacking the truthfulness property.

\section{Established Foundation}
\label{sec:foundation}
This section will summarize the formalisms of Carroll Structures, Carroll-aligned combinatorial LMSR and associated algorithms.
In addition to the code which we've provided, this section is the main source of technical documentation for the work of Fall 2025.

\subsection{Carroll Structures}

A {\em Carroll Structure} is a networked set $\Pi$ of propositions, where a proposition $\pi\in\Pi$ is either
\begin{itemize}
    \item An atomic proposition $\pi= p$, or
    \item An ordered pair of distinct propositions $\pi = (p',p''),$ such that $p'\neq p'', \pi\neq p'',$ and $ p'\mbox{ is atomic}.$
\end{itemize}
Thus, $\Pi$ is potentially recursive in the sense that a paired proposition can contain up to one other paired proposition; see $t$ in this example:
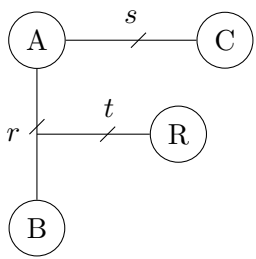
\begin{center}
    \begin{tikzpicture}[node distance=2.5cm,
        every node/.style={draw, circle}]
        \node (B) {B};
        \node (A) [above of=B] {A};
        \node (C) [right of=A] {C};
    
        \coordinate (M) at ($(B)!0.5!(A)$);
    
        \node (R) [right=1.5cm of M] {R};
    
        \draw (B) -- node[left, draw=none] {$r$} (A);
    
        \draw (M) ++(-0.1, 0) -- ++(0.2, 0.2);
    
        \draw (A) -- node[above, draw=none] {$s$} (C);
    
        \draw ($(A)!0.5!(C)$) ++(0, -0.1) -- ++(0.2, 0.2);
    
        \draw (R) -- node[pos=0.5, xshift=6pt, yshift=10pt, draw=none] {$t$} (M);
    
        \draw ($(R)!0.5!(M)$) ++(-0.1, -0.1) -- ++(0.2, 0.2);
    \end{tikzpicture}
\end{center}
The set of propositions here is
\begin{align*}
  \Pi=  \{  &A,\\
        &B,\\
        &C,\\
        &R,\\
        &r = (B,A),\\
        &s = (C,A),\\
        &t = (R,r)\}.
\end{align*}

The existence of a paired proposition $r=(B,A)$ enforces a logical constraint on the truth values of $r,B,$ and $A.$
This logical constraint is configurable, but in this document we treat the default constraint as the NAND of all three: $\neg (A\land B\land r)$; i.e., at most 2 of the propositions $r,B,A$ can ever be True simultaneously.
In Section~\ref{ssec: relationships}, we discuss the technical details of encoding other logical constraints besides NAND.
Note that we often refer to a paired proposition as an \textit{edge.}

\subsection{Combinatorial LMSR}

A logarithmic market scoring rule (LMSR) is a mathematical formalism which facilitates automated market making in the context of prediction markets.
The LMSR mechanism maintains a self-consistent probability distribution over a set of possible outcomes, where the probability of each outcome is signaled to traders in the form of a marginal price.
Mathematically, a basic LMSR works as follows~\cite{Hanson2003,Hanson2007}.
There is a set of outcomes $\Omega$; the LMSR allows traders to buy any security $\omega\in\Omega$.
Security $\omega$ is a contract of the form ``pays \$1 if outcome $\omega$ occurs, pays \$0 otherwise.''
Note that in this formalism, the outcome space is disjoint and complete: exactly one of the outcomes in $\Omega$ will occur.
Following~\cite{Chen2008a}, we write $\q=(q_\omega)_{\omega\in\Omega}$ to denote the number of outstanding shares for all securities.
The market is initialized with $\q=\q^0$; see Section~\ref{ssec: programmable markets} for specifics in our case.
The market maker tracks all outstanding shares $\q$ as time progresses using a cost function
\begin{equation}
\label{eq: basic cost function}
C(\q) = b\log \sum\limits_{\omega \in \Omega} e^{q_\omega/b}.
\end{equation}
Here, $b$ is a positive parameter called the \emph{liquidity parameter}; large values of $b$ result in larger amounts of capital being required to move the market prices.
The cost function acts as a potential function for trading: $C(\q)$ represents the amount of capital ``stored in'' the market maker, so that $C(\q^0)$ represents the market maker's initial subsidy and maximum eventual cost.%
\footnote{Note however that if any security $\omega\in\Omega$ has $q^0_\omega<0$, it can happen that the net cost of running the market maker can exceed $C(\q^0)$.}
Thus, the cost of a transaction which changes the number of outstanding shares from $\q$ to $\tilde{\q}$ is simply $C(\tilde{\q})-C(\q)$.

From a trader's standpoint, the instantaneous prices of the securities are interpreted as implied probabilities of the associated outcomes.
Due to the functioning of the market maker mechanism, the instantaneous price $p_\omega(\q)$ of security $\omega$ is simply given by the partial derivative of the cost function with respect to $q_\omega$:
\begin{equation}
	\label{eq: basic prices}
	p_\omega(\q) = \frac{e^{q_\omega/b}}{\sum\limits_{\omega' \in \Omega} e^{q_{\omega'}/b}}.
\end{equation}

A \textit{combinatorial} LMSR admits an additional layer of reasoning by allowing traders to make specific bets regarding logical combinations of events (e.g., if the outcome space is disjoint events $(A,B,C)$, a combinatorial LMSR allows trading in \emph{compound} securities such as $(A\ {\tt or}\ B)$).
A compound security $S\subset \Omega$ is a contract of the form ``pays \$1 if one of the outcomes in the set $S$ occurs, and pays \$0 otherwise.''
When a trader purchases $q$ shares of the compound security $S$, this is mathematically equivalent to purchasing $q$ shares of \emph{each} security $\omega\in S$.
We write $\Theta$ to denote the set of all compound securities, and $Q=(q_S)_{S\in\Theta}$ denotes the outstanding shares of all compound securities.
Noting that $q_\omega = \sum_{S\in\Theta:\omega\in S}q_S$, the cost and price functions in the combinatorial case take the same form as in~\eqref{eq: basic cost function} and~\eqref{eq: basic prices}, respectively.
Specifically, the price of a compound security $S$ is the sum of the prices of its members $\omega$:
\begin{equation}
\label{eq: combinatorial prices}
	p_S(Q) = \frac{\sum\limits_{\omega \in S}e^{q_\omega/b}}{\sum\limits_{\omega' \in \Omega} e^{q_{\omega'}/b}}.
\end{equation}

\subsection{Edges: How to encode relationships}
\label{ssec: relationships}

In this section, we begin with the following simple Carroll structure:
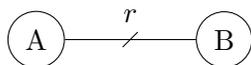
\begin{center}
	\begin{tikzpicture}[node distance=2.5cm,
		every node/.style={draw, circle}]
		\node (B) {B};
		\node (A) [left of=B] {A};
		
		\draw (B) -- node[above, draw=none] {$r$} (A);
		
		\draw ($(B)!0.5!(A)$) ++(-0.1, -0.1) -- ++(0.2, 0.2);
	\end{tikzpicture}
\end{center}
This structure contains propositions $\Pi=\{A,B,r\}$, where $r\equiv \neg(A\land B)$.
There are a variety of ways to think about this Carroll structure; for example, one may think of $r$ as a ``switch'': if $r\equiv \mbox{False}$, then $A$ and $B$ are logicallly independent, but if $r\equiv \mbox{True}$, then only one of $A$ or $B$ can be True.
I.e., $r$ switches between two different truth tables for $A$ and $B$.
Another way to think of it is that the Carroll structure itself is equivalent to the proposition $\neg(A\land B\land r)$.
This last framing particularly facilitates encoding the Carroll structure directly in the outcome space of a combinatorial LMSR.

To do this, we associate an outcome $\omega$ to each of the feasible truth-value assignments allowed by the Carroll structure (i.e., the assignments which satisfy $\neg(A\land B\land r)$).
For the example $A,B,r$ structure above, there are 7 such feasible assignments.
It is interesting to note that without the Carroll constraint, there are 8 feasible outcomes; thus, encoding the Carroll logic directly in the outcome space does result in some (perhaps small) computational benefits compared with a ``vanilla'' combinatorial LMSR.

\subsubsection{Other Types of Edges/Relationships}
Most of the testing and experiments associated with this document used the NAND interpretation of a paired proposition.
However, any other logical relationship is possible.
Two specific examples are presented here.
Note that in all of these, the relationship itself is a basic proposition and is thus contestable: markets decide the strength of the NAND/Support/Equivalence relationship.
\begin{itemize}
	\item {\bf Support.} 
	A ``Support'' relationship is a directed relationship which encodes an implication from the source to the target.
	That is, in a Carroll structure with $r=(B,A)$, having $r=\mbox{True}$ enforces the constraint that $B\implies A$ or equivalently that $\neg B\lor A$.
	In the associated combinatorial LMSR, only outcomes which satisfy $(A \lor \neg B \lor \neg r)$ are allowed.
	Since there are 7 such outcomes, the Support relationship has the same computational complexity as the standard NAND relationship.
	\item {\bf Equivalence.}
	An ``Equivalence'' relationship is an undirected relationship which forces the truth values of the source and target to be the same.
	That is, in a Carroll structure with $r=(B,A)$, having $r=\mbox{True}$ enforces the constraint that $B\equiv A$ or equivalently that $(A\land B) \lor (\neg A \land \neg B)$.
	In the associated combinatorial LMSR, only outcomes which satisfy $$\neg r \lor (A\land B) \lor (\neg A \land \neg B)$$ are allowed.
	Note that there are only 6 such outcomes; thus, Equivalence is more computationally efficient than NAND and Support.	
	
	\item {\bf Beyond the Binary: Hyperedges.}
	In principle, there is nothing preventing the implementation of logical relationships which go beyond simple graphical structures; for example, one could establish a switchable 3-way NAND relationship between $A,B,C$ by introducing a proposition $r$ and then allowing only outcomes which satisfy
    $$\neg (r\land A\land B\land C).$$
    This is completely mathematically principled and possible and may have advantages in some circumstances.
    However, the computational complexity consequences of any such approach need to be carefully considered.
    Our Scaling Proposal (Section~\ref{ssec: scaling}) likely depends heavily on sparse graphical structures.
\end{itemize}

%

\subsection{Programmable Markets}
\label{ssec: programmable markets}

To unlock the full potential of Carroll mechanisms and combinatorial LMSR, traders need the ability to add new propositions with as few constraints as possible.
Ideally, a trader should be able to add a new basic or paired proposition at any time.
However, this is not an affordance which has been considered by the classical LMSR literature --- traditionally, it is assumed that the outcome space is pre-specified by a centralized market operator and that it does not change during the course of trading.
Accordingly, we must be mindful of several potential problems that could be caused by dynamic structure:
\begin{enumerate}
	\item To prevent spamming, introducing a new proposition should not be free.
	\item The prices of any existing securities should not be changed when a new proposition is introduced.
	\item The liquidation values of all traders' open positions should not be changed when a new proposition is introduced.
	\item The initial price of the new proposition should be predictable and ``make sense'' to a lay trader. 
    This may be a difficult property to achieve.
\end{enumerate}

We have done some preliminary work on this concept of \emph{programmable markets} in an attempt to address these concerns.
However, it appears that some straightforward approaches may have problems which are difficult to overcome.

Let us consider two distinct operations:
\begin{itemize}
	\item Adding a new atomic proposition, and
	\item Adding a new paired proposition.
\end{itemize}

We consider these one at a time.

\subsubsection{Adding Atomic Propositions}
\label{sssec: adding atomic}

First consider the case of adding an atomic proposition $S$ to an existing Carroll structure $\Pi$.
Intuitively, it appears reasonable that adding only an atomic proposition with no edges would immediately satisfy concerns 2 and 3 from above because of $S$'s probabilistic independence from all other propositions.
We show here that this is indeed the case.

Our convention when adding a new atomic proposition $S$ is to initialize the number of shares to $0$; i.e., $q_S=0$.
Before adding $S$, denote the set of LMSR outcomes by $\Omega$; after adding $S$, denote the set of LMSR outcomes by $\tilde{\Omega}$.
Note that adding $S$ exactly doubles the number of outcomes, since for each outcome $\omega\in\Omega$, there is one in $\tilde{\Omega}$ with $S=\mbox{True}$ and one with $S=\mbox{False}$.
We use the convention that if $\omega\in\Omega$, we write $\omega\in\tilde{\Omega}$ to denote the version of $\omega$ in which $S=\mbox{True}$, and $\tilde{\omega}\in\tilde{\Omega}$ to denote the version of $\omega$ in which $S=\mbox{False}$.

Adding the new proposition requires us to update the outstanding quantities of all previously-existing shares and initialize the quantities of newly-created shares.
Let $q_\omega$ denote the quantity of shares of outcome $\omega$ before adding $S$, and let $\tilde{q}_\omega$ denote the quantity of these shares after adding $S$. 
For any $\omega'\in\tilde{\Omega}$ such that $\omega'\notin\Omega$, we likewise write $\tilde{q}_{\omega'}$ to denote the quantity of newly-created shares.

Because $\tilde{\Theta} = \Theta\setminus S$ and $q_S=0$, it holds for any $\omega\in\Omega$ (that is, $\omega\in S$) that the updated quantity satisfies $\tilde{q}_\omega = q_\omega$:
\begin{align*}
	\tilde{q}_\omega 	&= \sum_{T\in\tilde{\Theta}:\omega\in T}q_T \\
						&= q_S + \sum_{T\in\tilde{\Theta}\setminus S:\omega\in T}q_T \\
						&= \sum_{T\in\Theta:\omega\in T}q_T  = q_\omega.
\end{align*}
Similarly, for any newly-created $\omega'\in\tilde{\Omega}$ such that $\omega'$ is the $S=\mbox{False}$ version of $\omega\in\Omega$ (that is, $\omega'\notin S$), it holds that $\tilde{q}_{\omega'} = q_\omega$: 
\begin{align*}
	\tilde{q}_{\omega'} &= \sum_{T\in\tilde{\Theta}:\omega'\in T}q_T \\
						&= \sum_{T\in\tilde{\Theta}\setminus S:{\omega'}\in T}q_T \\
						&= \sum_{T\in\Theta:\omega\in T}q_T  = q_\omega.
\end{align*}

This means that we have $\tilde{p}_S(\tilde{Q}) = 1/2$, comfortably satisfying Concern \#4.
This is because for every $\omega\in\Omega$, we have that $\tilde{q}_\omega = \tilde{q}_{\omega'}$ and exactly half the securities in $\tilde{\Omega}$ are also members of $S$:
\begin{align}
\sum_{\tau\in \tilde{\Omega}} e^{\tilde{q}_\tau/b} 	&= \sum_{\omega\in S} e^{\tilde{q}_\omega/b} + \sum_{\omega'\notin S} e^{\tilde{q}_{\omega'}/b} \\
													&= 2\sum_{\omega\in S} e^{\tilde{q}_\omega/b},
\end{align}
implying that
\begin{align}
	\tilde{p}_S(\tilde{Q}) 	&= \frac{\sum\limits_{\omega \in S}e^{\tilde{q}_\omega/b}}{\sum\limits_{\tau\in \tilde{\Omega}} e^{\tilde{q}_\tau/b}} \\
							&= \frac{\sum\limits_{\omega \in S}e^{\tilde{q}_\omega/b}}{2\sum_{\omega\in S} e^{\tilde{q}_\omega/b}} = \frac{1}{2}.
\end{align}

Furthermore, for every other security $T\in\Theta$, $\tilde{p}_T(\tilde{Q}) = p_T(Q)$, which satisfies our original concern \#2.
This is because $\omega\in S$ is equivalent to $\omega \in \Omega$, and also $\tilde{q}_\omega = \tilde{q}_{\omega'}$:
\begin{align}
	\tilde{p}_T(\tilde{Q}) 	&= \frac{\sum\limits_{\omega \in T}e^{\tilde{q}_\omega/b}}{\sum\limits_{\tau\in \tilde{\Omega}} e^{\tilde{q}_\tau/b}}  \label{al2}\\
							&= \frac{\sum\limits_{\omega\in S,\omega\in T}e^{\tilde{q}_\omega/b} + \sum\limits_{\omega'\notin S,\omega'\in T}e^{\tilde{q}_{\omega'}/b}}{\sum\limits_{\tau\in S}e^{\tilde{q}_\tau/b} + \sum\limits_{\tau'\notin S}e^{\tilde{q}_{\tau'}/b}} \\
							&= \frac{2\sum\limits_{\omega\in S,\omega\in T}e^{\tilde{q}_\omega/b}}{2\sum\limits_{\tau\in S}e^{\tilde{q}_\tau/b}} \\
							&= \frac{\sum\limits_{\omega\in T}e^{\tilde{q}_\omega/b}}{\sum\limits_{\tau\in \Omega}e^{\tilde{q}_\tau/b}} \\
							&= \frac{\sum\limits_{\omega \in T}e^{{q}_\omega/b}}{\sum\limits_{\tau\in {\Omega}} e^{{q}_\tau/b}} = p_T(Q).  \label{al3}
\end{align}

Next, we need to verify that adding a new atomic proposition does not change the liquidation values of any outstanding shares (concern \#3); that is, the value of existing traders' positions should not be affected by adding a new proposition.
To do this, consider two arbitrary share quantity vectors for $\Theta$ denoted by $Q^1$ and $Q^2$.
For $i\in\{1,2\}$, write $\tilde{Q}^i$ to denote a quantity vector for $\tilde{\Theta}$ with $\tilde{Q}_T^i=Q_T^i$ for any $T\in\Theta$, and $\tilde{Q}_S^i=0$. 
If the cost functions are $C(\cdot)$ and $\tilde{C}(\cdot)$ (before and after adding proposition $S$, respectively), we require that 
\begin{equation}
	 \tilde{C}(\tilde{Q}^1) - \tilde{C}(\tilde{Q}^2) = C(Q^1) - C(Q^2).
\end{equation}

\begin{align}
	\tilde{C}(\tilde{Q}^1) - \tilde{C}(\tilde{Q}^2)	&= b\log \sum\limits_{\omega \in \tilde{\Omega}} e^{\tilde{q}^1_{\omega}/b} - b\log \sum\limits_{\omega \in \tilde{\Omega}} e^{\tilde{q}^2_{\omega}/b} \label{al0}\\
													&= b\log \frac{ \sum\limits_{\omega \in \tilde{\Omega}} e^{\tilde{q}^1_{\omega}/b}}{  \sum\limits_{\omega \in \tilde{\Omega}} e^{\tilde{q}^2_{\omega}/b}} \\
													&= b\log \frac{ \sum\limits_{\omega \in S} e^{\tilde{q}^1_{\omega}/b} +  \sum\limits_{\omega \notin S} e^{\tilde{q}^1_{\omega}/b}}{  \sum\limits_{\omega \in S} e^{\tilde{q}^2_{\omega}/b} +  \sum\limits_{\omega \notin S} e^{\tilde{q}^2_{\omega}/b}} \\
													&= b\log \frac{ 2 \sum\limits_{\omega \in S} e^{\tilde{q}^1_{\omega}/b}}{ 2 \sum\limits_{\omega \in S} e^{\tilde{q}^2_{\omega}/b}} \\
													&= b\log \frac{ \sum\limits_{\omega \in \Omega} e^{q^1_{\omega}/b}}{ \sum\limits_{\omega \in \Omega} e^{q^2_{\omega}/b}} = C(Q^1) - C(Q^2). \label{al1}
\end{align}
Thus, concern \#3 is satisfied.

Finally, we will address the pricing issue and show that the cost function implies that every new proposition added is charged the predictable and constant fee of $b\log 2$.
We will directly compute the quantity $\tilde{C}(\tilde{Q}) - C(Q)$.
We skip steps in the derivation which follow the logic of~\eqref{al0}-\eqref{al1}.
\begin{align}
	\tilde{C}(\tilde{Q}) - C(Q)	&= b\log \sum\limits_{\omega \in \tilde{\Omega}} e^{\tilde{q}_{\omega}/b} - b\log \sum\limits_{\omega \in {\Omega}} e^{{q}_{\omega}/b} \\
								&= b\log \frac{ 2\sum\limits_{\omega \in {\Omega}} e^{{q}_{\omega}/b}}{ \sum\limits_{\omega \in {\Omega}} e^{{q}_{\omega}/b}} = b\log 2.\\
\end{align}

\subsubsection{Adding Paired Propositions: Negative Initialization}
\label{sssec: adding paired}

For paired propositions (``edges''), it would be possible to ignore concerns \#2--\#4 and simply initialize an edge's shares at 0 in the same way as an atomic proposition.
This confers simplicity but also poor interpretability; edges would be initialized at an arbitrary-seeming price and affect other existing prices (and thus liquidation values) in an arbitrary-seeming way.
Another possible approach would be to follow the Golden-Ratio-Based method described in Section~\ref{sssec: golden ratio}.

For completeness, I also include here a concept which works within the market but fails upon market resolution.
A paired proposition $R$ could be added simply by initializing its security balance to something negative (e.g., $q^0_R=-10b$).
However, this often causes the market maker to be under-funded because it may dramatically increase the maximum payout of the market maker.
The maximum payout of the LMSR is not $C(q^0)$; rather, it is $C(q^0) + \sum_{i}(q_i^{0})^-$; by initializing $q_R^0$ to a negative number, you could end up with a giant un-funded piece due to the gap between the negative initialization and 0 on each edge.
One possible workaround would be to have the trader who adds an edge pay a (probably large) fee equal to the maximum payout shortfall.
In any case, it is instructive to include an analysis of the trading process with negative initializations.

In this case, a suitably large constant $K>>1$ is selected, the paired proposition $R$ is added and then $Kb$ phantom shares of $R$ are \emph{sold} (i.e., we initialize the negative balance of $q_R=-Kb$).
This ensures that all existing liquidation values and prices are unaffected by adding the new edge because the edge is added in the ``off'' state.
The effect of this is that the cost and initial price of any paired proposition is close to 0.

To see that this works, one simply needs to trace the arguments from Section~\ref{sssec: adding atomic} but with $q_R=-Kb$ to see that if $\omega\in R$, we have $\tilde{q}_\omega = q_\omega - Kb$ and if $\omega'\notin R$, we have $\tilde{q}_{\omega'} = q_{\omega'}$.
For example, the initial price of $R$ is close to $0$:

\begin{align}
	\tilde{p}_R(\tilde{Q}) 	&= \frac{\sum\limits_{\omega \in R}e^{\tilde{q}_\omega/b}}{\sum\limits_{\tau\in \tilde{\Omega}} e^{\tilde{q}_\tau/b}} \\
							&= \frac{\sum\limits_{\omega \in R}e^{(q_\omega-Kb)/b}}{\sum\limits_{\omega \in R}e^{(q_\omega-Kb)/b} + \sum\limits_{\omega' \notin R}e^{q_{\omega'}/b}} \underset{K\to\infty}{\longrightarrow} 0.
\end{align}
The other arguments analogous to~\eqref{al2}-\eqref{al3} and~\eqref{al0}-\eqref{al1} follow cleanly in a similar fashion.

\subsubsection{The Golden-Ratio-Based Method}
\label{sssec: golden ratio}
I have not had the time to verify analytically that this works (or why this works), but numerically I discovered a different way to initialize new points in the market which seems to be quite beautiful and \textit{should} avoid the insolvency of negative initializations.
Very briefly, it appears that all concerns (\#1---\#4) are satisfied if the graph is a tree and you do the following.
Suppose paired proposition $R=(A,B)$ is added to the Carroll Structure.
Let $q^*=\log \phi\approx 0.481$, where $\phi\approx1.6108$ is the golden ratio.
Then for each $S\in\{R,A,B\}$, increment the number of shares outstanding of $S$ by $q^*$; that is, $q_S +\hspace{-1.5mm}= q^*$. 
It appears that \textit{as long as the graph maintains a tree structure}, this results in new propositions being added with an initial price of $0.5$, costing about $1.92 b$, and otherwise satisfying concerns (\#1---\#4).
It is not clear how this interfaces with the scaling proposal in Section~\ref{ssec: scaling}.


\subsection{Scaling Proposal: (not fully tested or analyzed)}
\label{ssec: scaling}
One of the most significant challenges posed in the space of combinatorial LMSR work is that of computational complexity.
The problem of computing prices in combinatorial LMSR is well-known to be \#P-Complete~\cite{Chen2008a}, and despite our slight reductions relative to the upper bound of ${\cal O}(2^{|\Pi|})$ on the number of outcomes, a direct implementation of our system still scales exponentially.

However, our initial experiments suggest that tree-like (and likely any sparse) Carroll structures can be exploited to alleviate the scaling problem, possibly eliminating it entirely.
Our solution splits a large Carroll structure $\Pi$ into two similarly-sized parts $\Pi^1$ and $\Pi^2$ in such a way that there is exactly one linking proposition $L$ which is in both parts.
If a trader purchases (or sells) a security $A\in\Pi^1$, this causes a discrepancy between the price of the linking proposition $L$ in the two parts.
To reconcile this discrepancy, the system executes a \emph{phantom} purchase of $L$ in $\Pi^2$ to equalize the prices of $L$ in $\Pi^1$ and $\Pi^2$ (in the provided Python code, this is done by the {\tt Market\_Maker.buy\_to\_price} method).
Likewise, if a trader purchases (or sells) a security $B\in\Pi^1\setminus\Pi^2$, the system executes a phantom purchase of $L$ in $\Pi^1$ to equalize the two $L$ prices.

The scaling advantage here is huge, provided that there is a sufficiently small upper bound on the size of each part $\Pi^i$.
Suppose that at most $M$ propositions are allowed in each part.
Then the total number of outcomes is never more than about $|\Pi|/M\cdot2^M$ (since the Carroll structure is split into about $|\Pi|/M$ pieces); i.e., an asymptotic upper bound of ${\cal O}(|\Pi|)$ with a constant which is exponential in $M$.
If $M$ is kept low (i.e., 10 or less), real-world performance should be quite good.

We have not performed an in-depth mathematical analysis to probe the limits of this scaling approach, but our numerical experiments demonstrate that it constitutes a very close approximation to an exact implementation.
To study further, the following properties need to be verified for any sequence of trades and any partition of $\Pi$:
\begin{enumerate}
	\item The prices must be the same in the segmented Carroll structure as in the exact Carroll structure, and
	\item A trader must end with the same liquidation value remaining in the segmented Carroll structure  as in the exact Carroll structure.
\end{enumerate}

\subsubsection{Cyclic Carroll Structures}
In the case that the underlying Carroll structure contains cycles, we conjecture that the segmented approximation is no longer exact, but that if care be taken to choose an appropriate segmentation, the approximation error can be controlled.
Our reasoning is as follows: suppose there are two parts $\Pi^1$ and $\Pi^2$ which share two linking propositions $L$ and $L'$. 
If a purchase is made of security $A\in\Pi^1$, this affects the prices of $L$ and $L'$ in $\Pi^1$.
However, the prices of $L$ and $L'$ are also connected by some path in $\Pi^2$ --- thus, in the exact Carroll structure, there would be some additional price effects between them which are not exactly summarized in the $\Pi^1$ part.
Perhaps an iterative sequence of phantom purchases in $\Pi^1$ and $\Pi^2$ could resolve this, but this is not clear and we have no conjecture about whether or how quickly this iteration might converge.

However, we make two observations which likely can be used to control the approximation error:
\begin{enumerate}
	\item If the path in $\Pi^2$ connecting $L$ and $L'$ is very short, then it is likely that a re-segmentation could entirely include the offending cycle into one of the Carroll parts.
	In this case, the problem vanishes entirely.
	\item If the path in $\Pi^2$ connecting $L$ and $L'$ is very long (thus rendering it impossible to include the cycle into a single Carroll part), then it is likely that the price interaction between $L$ and $L'$ in $\Pi^2$ is very weak and can thus be ignored.
\end{enumerate}
Hopefully this pair of observations can be formalized somehow in a way that clearly parameterizes the approximation error.
Ideally, one would obtain a formal statement which qualitatively says something like ``either the approximation error is small or it is easy to fix.''

\subsubsection{Possible worst-case graphs to investigate}
\begin{itemize}
	\item Suppose that in the local Carroll part, the purchase in $\Pi^1$ causes $L$ to go high and $L'$ to go low, but $L$ and $L'$ are directly connected by a heavily-supported Equivalence edge.
	That is, $L$ and $L'$ are simultaneously driven apart \textit{and} tied tightly together.
	In an exact Carroll structure, these two effects would dampen each other (and provide evidence \emph{against} the Equivalence edge); however, using our approximation technique directly, $L$ and $L'$ would be constrained to follow the $\Pi^1$ part and thus likely result in a large approximation error.
	The easy resolution would be to re-segment to include the equivalence edge in one part.
	However, it would be very interesting to investigate possible ways to mitigate without re-segmentation.
	For example, perhaps an acceptable approximation could be obtained by executing an iterative series of phantom trades on $L$ and $L'$ in both $\Pi^2$ \textit{and} $\Pi^1$.
    This iteration may converge in a similar way to Laplacian consensus~\cite{Saber2003,Olfati-Saber2007}, and perhaps similar analysis techniques can be used. 
	\item It is possible that this pathology could be made much worse by simply lengthening the chain which connects $L$ and $L'$ in $\Pi^2$, essentially emulating a single equivalence edge by chaining many strongly-supported equivalence edges.
	This would have the same effect as the first, but would be less amenable to re-segmentation
    If an iterative approach fixes the first, it would almost certainly also work here.
\end{itemize}

\section{Epistemic Leverage: Open Questions}
\label{sec: EL}
This section will focus on the concept of Epistemic Leverage (EL) in the form that we studied in this project.

\subsection{High-Level Desired Functionality}
\label{ssec: functionality}
Goals: 
\begin{itemize}
    \item What do we want this system to do?
    \item What affordances do we want this system to have?
\end{itemize}

The original goal of the epistemic leverage mechanism set 
is to give 
disproportionate influence to market participants known for their capacity to change their mind.
In this iteration of the work, we explored techniques which attempt to achieve this by providing incentives for participants to reveal the reasoning behind the positions they take.
In addition, several of our stated goals are to design a mechanism which can:
\begin{itemize}
    \item Reveal private information including private reasoning (i.e., our desired mechanism set would infer the state and relevance of latent variables),
    \item Give a true grounding to the mutual relevance of various claims,
    \item Give more influence to concrete (``low ambiguity'') statements, and
    \item Fund information provision such as research and journalism.
\end{itemize}

\subsubsection{Basic Affordances}

The basic proposed  mechanism set includes the following primitive affordances:
\begin{itemize}
    \item {\bf Restake:} An agent can say why they believe a point in exchange for additional influence on that point. 
    We have typically envisioned this coming in the form of a statement like ``I believe $A$ is True, but I would change my mind if I discovered that $B$ were True.'' 
    This can be interpreted in a couple ways, but we've tended to think of it as a purchase of $A$ which also gives some signal to $r$ (or a purchase of $r$ which may provide some bonus signal to $A$).
    See the following diagram:
    
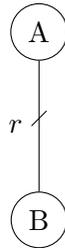
\begin{center}
        \begin{tikzpicture}[node distance=2.5cm,
            every node/.style={draw, circle}]
            \node (B) {B};
            \node (A) [above of=B] {A};
        
            \draw (B) -- node[left, draw=none] {$r$} (A);
        
            \draw ($(B)!0.5!(A)$) ++(-0.1, 0) -- ++(0.2, 0.2);
        \end{tikzpicture}
    \end{center}
    Alternatively, there is a potential implementation in which the trader purchases a restake contract which then gradually ``evaporates'' away; the evaporation rate depends on market conditions, doubt from other participants, and possibly other aspects of market state.
    The primary purpose of restake is to reward transparent reasoning; a secondary effect may be that it elicits some information about relevance edges.
    All alone, restake may incentivize spurious or uninteresting/uninformative claims of relevance; see Section~\ref{ssec: requirements}.
    Thus, it needs to be designed carefully so as to avoid this mis-incentivization.

    In summary, this ``careful design'' requires that Restake can only provide a meaningful benefit if it is executed when $B$ is high (and $r$ is low) and then $B$ falls significantly (and $r$ rises significantly); for more detailed reasoning, see Section~\ref{req: double vindication}.
    
    \item {\bf Doubt:} Doubt allows other traders to call a restaker's bluff in order to act as a deterrent to a certain form of insincere restaking.\footnote{If we had resolvable edges, we probably wouldn't need a deterrent; perhaps it's useful to think of doubt as a component of an edge-resolution mechanism.}
    The core use-case of doubt is that a restaker has stated ``I like $A$, but I'd change my mind if $B$ were true. 
    Right now $B$ is high but I'm betting it will drop. 
    If it stays high, I'll drop my attachment to $A$.''
    A doubter buys a contract which says ``I bet you won't actually change your mind if $B$ stays high.''
    Then if $B$ stays high and the restaker \emph{does} change their mind, the doubter gets burnt.\footnote{The idea here is that if you're a person who is believed to be likely to change your mind (when you're shown to be wrong), you're unlikely to be doubted. 
    This effect empowers people who are known to be intellectually honest.}
    If $B$ stays high and the restaker \emph{doesn't} change their mind, the doubt contract entitles the doubter to a trickle of the restaker's extra influence.

    \item {\bf Mindchange:} If Restake means ``I'd change my mind if $B$ stays high'', then Mindchange is the follow-through where I actually reposition myself over a relevance edge.
    One hoped-for benefit of the Mindchange operation is that it \emph{may} give relevance information: if I've said ``I believe $A\land r$'' and then switch to $B \land r$, this may be a sign that I really did mean it.
    This may be asymmetric in the sense that it gives more information if the market moves against the restaker than otherwise.
    
\end{itemize}
    
\subsubsection{Desired Properties}
\label{ssec: properties}
We want the Carroll Mechanism set to be better than incumbent systems (such as Futarchy, token voting, representative democracy) in the sense that a well-informed and sincere population of traders can more easily counteract the bullying effects of money, and that the reasoning behind decisions can be explicitly adjudicated in-mechanism.
Two properties which we specifically want are the following:

\begin{itemize}
    \item {\bf Relevance Elicitation:} Ordinary decoupled prediction markets can only discover relevance in the negative. 
    If we have the structure {\tt B--r--A}, and $A$ and $B$ both resolve to True, then it is thus discovered that $r$ is False.
    If neither $A$ nor $B$ are true, then perhaps it does not matter much.
    But if one of $A$ or $B$ is true, relevance logically constrains these to act as mutual negations of each other.
    In this case, we would really like to know if $A$ and $B$ are believed to be mutually exclusive.
    
    \item {\bf Sincerity Advantage:} all else equal, the stake of a sincere trader should ultimately have more weight than that of an insincere trader.
    This may be enhanced by some kind of reputation effects, and may require a trader to be \emph{believed to be sincere} to work.
    This distinction may actually be one of the key research questions we need to address!
    See Sec.~\ref{question: reputation}.

\end{itemize}

\subsection{Mechanism Requirements}
\label{ssec: requirements}
This section will detail some of the requirements we have on the various mechanisms and affordances.
Several of the requirements are motivated by specific attack/manipulation models.

\subsubsection{Requirement: Reward for sincere restake}
\label{req: sincere restake}
{\bf Consequence: restake provides additional signal to $A$.}\\
This is the central functionality of the Restake operation; traders should be rewarded for revealing relevance information.
If a trader restakes $A/r/B$, this means they believe $A$ provided it also holds that $\neg B$.
This influence \emph{perhaps} may not come at the moment of restake: it may be that the additional influence is awarded over time, or only once the bets are seen to be successful.
Note also that this is a very broad requirement; the following subsections will pare it down and give additional bounds on how it can be done.

\subsubsection{Requirement: Restake Rewards Double Vindication}
\label{req: double vindication}
{\bf Consequence: if a restake is executed when $B=1$ and $r=0$ and then both $B$ and $r$ change values, then restake provides a substantial bonus.}
This is the golden use-case for Restake: the trader makes a doubly unpopular, correct bet and then gets it right on both sides.
If there is any setting in which restake gives a big bonus, it's here.
Note that this is an implication with an AND in the hypothesis; that is, it's very narrowly scoped:
\begin{quote}
If $B$ falls and $r$ rises, then restake provides a substantial bonus.
\end{quote}
This does \textit{not} necessarily mean that restake provides any bonus if $B$ falls and $r$ is constant, or if $r$ rises and $B$ is constant.

\subsubsection{Requirement: No Reward for Spurious Information}
\label{req: spurious restake}
{\bf Consequence: restake can only help $A$ if the counterpoint $B$ falls.}\\
This basic idea is that restake should not offer any special advantages to a trader who restakes on a proposition that is \emph{plainly false.}
Here are three relevant attacks; note in each of these the attacker creates the counterpoint \textit{for the purpose of insincerely restaking on it}:
\begin{enumerate}
	
	\item{\bf Uninformative claims of relevance.}
	\label{failure mode: uninformative restake}
	I could say ``I think we should launch the satellite, but I would change my mind if it were proved that the Earth were flat.''
	I.e., I've restaked on a \textit{relevant} claim that I'm sure will resolve False.
	
	\item{\bf Spurious claims of relevance.}
	\label{failure mode: spurious restake}
	I could say ``I think we should launch the satellite, but I would change my mind if the freezing point of water were -10 degrees Celsius.''
	This is even worse than uninformative, since it's also insincere about the relevance: I know the freezing point of water is 0, but I actually wouldn't change my mind, even if it weren't. 
	But I'm safe from being caught out in my insincerity, because I know the freezing point of water isn't -10.
	

	\item{\bf Claims of ambiguous relevance.}
	\label{failure mode: ambiguous relevance}
	I could say ``I think we should launch the satellite, but I would change my mind if there were life on Europa.''
	This statement is ambiguous by design: we don't know if there is life on Europa; even if we did, it's not clear whether that would contraindicate launching the satellite.
	We might call both the counterpoint and the relevance \textit{plainly ambiguous,} and a restake mechanism should not reward restaking in this scenario.
\end{enumerate}

Note that these examples of potential attacks cannot be mitigated by a doubt-like mechanic because spurious restake is not falsifiable when the counterpoint is plainly false or ambiguous.
The unifying feature is that the signal on $B$ is constant throughout the interaction.
That is, regardless of what the relevance signal on $r$ does, restake should not give any special advantages in cases when $B$'s signal does not change.
This implies the following:
\begin{proposition}
\label{prop: decreasing in B}
The advantage of an $A/r/B$ restake (relative to buying $A$ directly) is nonincreasing in $\Delta B$, where $\Delta B$ is the change in the price/signal of $B$, and restake bonus is $\leq0$ when $\Delta B \geq 0$.
\end{proposition}
\noindent This has some connection to the Asymmetric Sincerity Verification challenge: see Section~\ref{challenge: asymmetric sincerity}.

\subsubsection{Requirement: No Reward for Irrelevant Restake}
\label{req: irrelevant restake proof}
{\bf Consequence: Restake can only help $A$ if the relevance $r$ rises.}\\
An attacker might try to get a restake bonus when an irrelevant $B$ decreases despite the favor on $r$ never being high.
The threat comes in two forms:
\label{threat: irrelevant restake}
\begin{enumerate}
	\item {\bf Spurious Edge Variation:} Alice is an $A$-proponent and knows of a false proposition $B$ that's currently over-valued.
	Alice links $B$ to $A$ with a new edge $r$ and restakes $A/r/B$; when the favor of $B$ falls, Alice gains a restake bonus on $A$.
	
	\item {\bf Irrelevant $B$ Variation:} At a higher cost, this could be executed as a variant of attack~\ref{threat: manipulation for restake} where Alice instantiates and intentionally over-values an irrelevant $B$ and then lets it crash.
	The reference implementation initiates the price of $B$ at $1/2$, so this would be an important concern (though a fix for RIA Variation 3 should handle this case, see Section~\ref{req: restake initialization proof}).
\end{enumerate}
In this attack, if markets are wise in the end, prices should eventually go to $B=r=0$.
The cost of the attack may be low depending on the implementation details in Programmable Markets.
%
%
%
One simple way to satisfy this requirement is to ensure that the restake benefit is simply increasing in $r$, which leads to the following which mirrors Proposition~\ref{prop: decreasing in B}:
\begin{proposition}
	\label{prop: increasing in r}
	The advantage of an $A/r/B$ restake (relative to buying $A$ directly) is nondecreasing in $\Delta r$, where $\Delta r$ is the change in the price/signal of $r$, and restake bonus is $\leq0$ when $\Delta r \leq 0$.
\end{proposition}
Note: Proposition~\ref{prop: increasing in r} is not saying that the restake bonus needs to be a function only of $\Delta r$.
For instance, one could imagine the restake bonus taking 2 arguments: (1) $\Delta r$, and (2) $r_{\rm initial}$, the price of $r$ at the moment of restake.
The maximum possible bonus might be decreasing in $r_{\rm initial}$, and the realized bonus would then be increasing in $\Delta r$.

\subsubsection{Requirement: Restake-Initialization-Attack (RIA)-Proof}
\label{req: restake initialization proof}
{\bf Consequence: upper bounds on the restake bonus, or lower bounds on the cost of restaking.}\\
The Restake operation must not allow a trader to profitably manipulate a plainly-false (but relevant) $B$ to have a temporarily high price just to gain a restake bonus.
Considering properties~\ref{req: sincere restake} and~\ref{req: double vindication}, it is clear that in a state where $B$ is high and $r$ is low, an $A/r/B$ restake must provide some advantage to $A$ relative to simply buying $A$.
The potential threat here is the following family of attacks:
\begin{enumerate}
	\item 
	\label{threat: manipulation for restake}
	{\bf Basic Attack:} $A$-motivated player Alice does the following in quick succesion: (1) establishes a ``plainly false'' but plainly relevant proposition $B$ such as ``The freezing point of water is -10 degrees,'' (2) links $A$ to $B$ via edge $r$, (3) purchases a large quantity of $B$ (perhaps using a sockpuppet), and (4) restakes $A/r/B$.
	Since the price of $B$ was high (and $r$ was low) when Alice restaked, Properties~\ref{req: sincere restake} and~\ref{req: double vindication} give her a big boost to $A$ when $B$ eventually falls.
	Alice then sells her stake in $B$ and recovers whatever of that capital remains.
	\item 
	{\bf Pre-existing $B$ Variation:} 
	Alice may be able to find an existing proposition $B$, link it, and boost it, rather than establishing one of her own. 
	This skips the costly step (1) in the Basic Attack.
	\textbf{Pro:} This may save Alice the cost of creating $B$. 
	\textbf{Con:} It may be difficult to find an existing False+Relevant proposition.
	\item 
	{\bf Manipulation-Free Variation:}
	Alice may skip step (3), avoiding the manipulation; this lowers the cost of the attack.
	Whether this works will depend on the specific model for adding new propositions to the Carroll structure; specifically, it depends on the initial prices for newly-added propositions as discussed in Section~\ref{ssec: programmable markets}.
\end{enumerate}
At least the Basic Attack and the Manipulation-free variation are always executable because it's easy to create relevant but false statements about anything you might want to do.
``The Earth is flat'' (relevant to anything involving astronomy) and ``The freezing point of water is -10'' (relevant to anything involving plumbing) are straightforward examples.
Thus, this attack can also be spammed.
By~\ref{req: double vindication}, Restake is required to give a big bonus in this exact scenario (high $B$, low $r$ and then both swap).
Since this attack cannot be prevented structurally, it must be disincentivized by ensuring that it's always unprofitable compared to some other ``pro-social'' action (such as purchasing $A$ directly).

The cost of this attack depends on which variation is executed; at most (variation 1), it is equal to the loss experienced by the attacker on shares in $B$ purchased and then resold plus the cost of creating proposition $B$.
It must be better for the attacker to invest this cost directly in $A$, rather than to execute the attack and get the restake bonus. 
{\bf Thus, this requirement implies a family of upper bounds on the bonus achievable from restake, or a family of lower bounds on the cost of entering a restake contract.}

Note that the cleanest version of this bound (coming from the Manipulation-Free Variation) says to compare these 2 options:
\begin{enumerate}
	\item Create propositions $B$ and $r$ (in one proposed implementation, this has cost $b\log 2$) and restake $A/r/B$ to gain restake advantage $\tt RES$.
	\item Invest $b\log 2$ capital in proposition $A$, to gain signal advantage $\tt SIG$.
\end{enumerate}
Then RIA-Proofness requires that $\tt RES\leq SIG$.
In other words, the restake bonus on $A$ associated with $\Delta B = -0.5$ and $\Delta r = 1$ must be less than the advantage to $A$ when $b\log 2$ capital is invested directly in $A$.

\subsection{The Mechanism Challenge}

Here, we discuss some salient challenges of implementing the mechanism; for example, apparent conflicts between the use-cases of various affordances.

\subsubsection{Are Doubt and Mindchange Necessary?}
\label{ssec: doubt and mindchange mechanism}
The discussion in Section~\ref{ssec: requirements} makes it clear that an $A/r/B$ restake can only provide advantages if the restake is executed when $B$ is high, and then the signal on $B$ drops.
Thus, the design use-case for Doubt is this: $B$ is high, and then either $B$ stays high or drops; the doubter is betting that the restaker will change their mind \textit{in the event that $B$ stays high,} but there's no particular bet in the event that $B$ drops.

In our approaches where there is an explicit restake bonus awarded to the restake, we often think of the doubter as purchasing a trickle of that restake bonus.
However, our core understanding of restake is that restake can only help $A$ if $B$ falls (See Section~\ref{req: spurious restake}).
That is, if $B$ stays high, there is no restake bonus for the doubter to earn!
So think through it from a different standpoint: Restake is a contract which says ``if $B$ drops and $r$ rises by the deadline, you get a bonus to $A$.'' 
The restaker pays a contract fee to enter into this contract. 
Doubt is a contract which entitles the doubter to a trickle of that fee. 
Likewise, the doubter pays a doubt premium to enter into this contract. 
Here are the possibilities:
\begin{itemize}
    \item $B$ stays high: 
    \begin{itemize}
        \item If the restaker does \textbf{not} mindchange, then the doubter is vindicated.
        \begin{itemize}
            \item Eventually the doubter earns the whole restake fee, and
            \item The doubter gets their premium back.
        \end{itemize}
        \item If the restaker \emph{does} mindchange, then:
        \begin{itemize}
            \item The doubter loses their premium, perhaps somehow it funds the restaker's reposition to $B$.
        \end{itemize}
    \end{itemize}
    \item $B$ drops low (the restaker is vindicated): 
    \begin{itemize}
        \item If the restaker does \textbf{not} mindchange,
        \begin{itemize}
            \item The doubter gets their premium back, perhaps subject to a fee.
            \item The sooner $B$'s signal falls, the more of the restake fee still remains.
        \end{itemize}
        \item If the restaker \emph{does} mindchange, then:
        \begin{itemize}
            \item The doubter loses their premium, perhaps somehow it funds the restaker's reposition to $B$.
            \item This is a weird outcome: it's unclear why the restaker would mindchange in this scenario.
        \end{itemize}
    \end{itemize}
\end{itemize}

This might all work --- but the value-add of the Doubt affordance is unclear.
It appears that Restake is a ``small'' enough affordance (i.e., applicable in a small enough set of circumstances) that the only cases where Restake is effective ($B$ drops and $r$ rises) are cases which the mechanism can adjudicate on its own.

\subsubsection{What does Mindchange Mean Epistemically?}
\label{sssec: mindchange epistemics}
We have typically described the mindchange mechanic as a way for a player to abandon something they used to want in response to new information.
As we have thought about this, the typical story has been something like ``I believe $A$, but if new information came to light about $B$, I would abandon my belief in $A$ and switch to $B$.''
This \textit{could} possibly be interpreted as something a Bayesian agent might do, but that isn't really how we think about it.
We really discuss this type of thing more as though the agent is changing priors.
This highlights a gap: we need to be much clearer about how we want to model epistemics in future iterations of this project.
I have noted that some study in epistemic game theory should be prioritized in future research (Section~\ref{sssec: lit review}).

In any case, we need to be very careful that \textit{if} we continue using the LMSR framework, we ensure that operations like Mindchange have a clear meaning in terms of a changing posterior or changing prior belief about the world.
Otherwise, their price effects may actually not be consistent with what we claim they are about.

\subsubsection{Asymmetric Sincerity Verification}
\label{challenge: asymmetric sincerity}

One of the key difficulties here is that the ``sincerity verification'' process is asymmetric; this is connected to questions of reputation in Section~\ref{question: reputation}.
Suppose that \textit{insincere} means ``I'm lying when I say I'd change my mind if $B$ turned out to be true,'' then we have 2 cases:
\begin{itemize}
    \item The player \textit{doesn't} change their mind when $B$ turns out to be True. 
    In this case, we can just look and see whether they changed their mind. 
    \item The player {doesn't} change their mind when $B$ turns out to be False. 
    Would they have changed their mind if $B$ had gone True?
    We have no way of knowing, so we can't check their sincerity in this case.
    {\bf Somehow, the mechanism had better \textit{not} reward this behavior.}
\end{itemize}

\subsubsection{How do we empower good without empowering bad?}

This challenge boils down to something like this:
\begin{quote}
{\bf Is it possible to give powerful tools to the ``good guys'' without also letting them be used by ``bad guys?''}
I.e., suppose we came up with a nice mechanism set that favors sincere players.
How can we be sure that wealthy insincere players can't simply impersonate sincere players and gain the same advantages for themselves?
\end{quote}

It appears clear that basic LMSR (and many other market scoring rules as well, see~\cite{Othman2010,Chen2011}) is subject to a variety of manipulation effects and attacks when used as the basis of a decision market.
We should therefore expect that any pure-LMSR-based approach would inherit these vulnerabilities, and thus that advantages must come from some extra-LMSR affordances (such as reputation effects).
If this is the case, those extra-LMSR affordances may disrupt some of the key guarantees we thought we were getting from LMSR.
In any case, care must be taken.



\subsection{Failure Modes and Threat Models}
\label{ssec: failure modes}
In this section, we detail some additional ways in which an otherwise-functional system might promote undesired behavior, either by providing inadvertent intrinsic incentives or by enabling a self-interested (or even malicious) attacker to subvert the mechanism.

\subsubsection{Risk: Restake provides too strong a negation effect.}
\label{failure mode: restake negates}
    This depends significantly on how restake is implemented.
    However, the potential issue here is that restaking $A/r$ may strongly force $B$ down. 
    I am not aware of a scenario where it fails to do this, since it has to push $r$ up and then typically at least protects $A$ if not boosts $A$.
    Those moves are intrinsically anti-$B$ and some implementations of restake basically amount to cheap $B$-negation (for instance, if retsake is interpreted as purchasing $A\land r$).
    The issue is that if you are effectively allowed to negate $B$, what stops you from using that power to negate $A$?
    Just imagine a simple setting like this: 
    
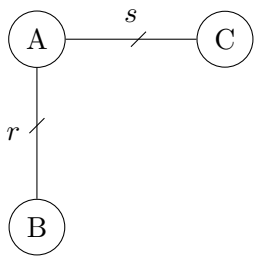
\begin{center}
        \begin{tikzpicture}[node distance=2.5cm,
        every node/.style={draw, circle}]
        \node (B) {B};
        \node (A) [above of=B] {A};
        \node (C) [right of=A] {C};
    
        \draw (B) -- node[left, draw=none] {$r$} (A);
    
        \draw ($(B)!0.5!(A)$) ++(-0.1, 0) -- ++(0.2, 0.2);
    
        \draw (A) -- node[above, draw=none] {$s$} (C);
    
        \draw ($(A)!0.5!(C)$) ++(0, -0.1) -- ++(0.2, 0.2);
        \end{tikzpicture}
    \end{center}
    I'm a wealthy opponent of $A$.
    What I can do is add a point $C$ meaning something obviously true like ``the freezing point of water is 0 degrees Celsius.''
    Then I can restake $C/s$ in an attack on $A$.
    It would be a sincere defense to object to $s$, but there is nothing here to intrinsically break the symmetry between the sincere and insincere traders.

\subsubsection{Capital Extraction Attack}
\label{sssec: capital extraction}
    Any implemented mechanism has to prevent a trader and their sockpuppets (Alice and ``Balice'') from executing a series of trades which enter and then exit and result in the trader profiting.
    If this were possible, a sophisticated (and perhaps patient) trader could repeat the exploit over and over and drain liquidity from the mechanism while providing no meaningful information.

    These kinds of attacks are fundamentally impossible if the entire mechanism set is simply composed of LMSR transactions.
    However, epistemic leverage likely will require something in addition to basic LMSR transactions, and here we need to be careful: both of our implementations of something like EL allowed capital extraction once we dug deep enough (short-selling~\ref{ssec: EL as short selling} and phantom shares~\ref{ssec: EL as phantom shares}).

    Several times, we've come up with some kind of epistemic leverage mechanism, only to have it foiled by an agent with a sockpuppet account.
    Future research needs to include some study of a comprehensive way to check for this and prevent it.

\subsubsection{Doubt is easily exploited by monied interests to suppress restake.}
    There is a potential concern that the Doubt affordance could be used to steamroll sincere restakers and reduce Restake to nothing.
    If Doubt is included in a future mechanism set, this concern needs to be addressed.

\subsection{Implementation Attempts}
\label{ssec: impleentation attempts}

\subsubsection{Epistemic leverage as a leveraged purchase}
\label{ssec: EL as short selling}

The core idea here comes from the fact that an LMSR may allow a trader to ``short'' a security in a mathematically meaningful way: a trader can take a negative (short) position in a security in exchange for cash.
A short security $A$ is a contract in the form of ``owes \$1 if $A$ resolves to True.''
It may be possible to implement epistemic leverage using this primitive: a trader who holds $A$ can ``restake'' on $r$ by opening a short position in the security $\neg r$ (mathematically equivalent to going long on $r$, except in the form of debt) and then using the short sale proceeds to fund a purchase of more $A$.
The outstanding $A$ shares are locked by the mechanism and held as collateral to secure the loan of $\neg r$.
In my formulation, these short shares in $\neg r$ are never explicitly traded but are simply represented mathematically as a means to provide the desired leverage.
This exposes the restaker/shorter to more risk because it's leveraged, it pushes up $A$ and $r$, and pushes down $B$.

Because a short position could resolve as a liability to the trader, it would be crucial to ensure that a trader who opens a short position be forced to maintain sufficient collateral in the contract to cover this loan if the market resolves unfavorably to them.
To accomplish this, the mechanism would have some kind of ``margin call'' mechanic which seizes the locked $A$ capital to unwind the position in the event that the market moves against the trader.
There are a variety of possible ways to implement such a margin call mechanic, but we have been unsuccessful in identifying a mechanic which does not allow a form of capital extraction attack as in Section~\ref{sssec: capital extraction}.
It remains an open question whether a margin call mechanic exists which does not permit some form of capital extraction.
\notetoself{I have some good notes on this (see REALLY SERIOUS MAJOR ISSUE on the ``Algorithms for implementation'' note).}

\subsubsection{Epistemic leverage as a phantom purchase}
\label{ssec: EL as phantom shares}
We also had the idea that perhaps we could implement epistemic leverage as a series of phantom purchases, currency rebasing, and burning.
The basic idea was like this:
Alice holds shares in $A$.
If she then buys shares in $r$, this automatically triggers an $A/r$ restake.
Due to the NAND character of $r$, any purchase of $r$ intrinsically pushes \textit{down} the price of $A$.
That is, when Alice revealed information about $r$, this actually hurt her in her position of $A$; the restake mechanism serves to compensate her for this action.
To compensate her, we perform a \textit{phantom}%
\footnote{A phantom purchase in security $A$ is a simple increase in $q_A$ without crediting the newly-created shares to any trader.
Since this increases the value of the cost function without an intake of new capital, it either must be paid for by reducing some other balances in the system or by risking insolvency when the market resolves.}
purchase of shares of $A$ to push the price of $A$ back up to where it was before she bought $r$.
To fund that phantom purchase, we burn a sliver of every outstanding share balance, effectively socializing Alice's protection.
(Alternatively, it would potentially be possible to leave the protection un-funded, and potentially risk insolvency at the time of market resolution.)

Upon implementation, we immediately discovered a capital extraction attack (Section~\ref{sssec: capital extraction}) in which Alice's sockpuppet (``Balice'') begins with a large stake in $A$, then Alice purchases $r$, then Balice sells her $A$ (which are now worth more due to the phantom shares), then Alice sells her $r$ which is now worth a huge amount more.
The bulk of the profits actually come from Alice selling the $r$ (plus a little when Balice sells the $A$), but the net effect is that Alice can just show up with her sockpuppet and immediately extract a significant amount of capital from other traders.
It may be that the burning step is central to the attack; perhaps this could be circumvented by buying the phantom shares at a mechanism loss.
However, it seems more and more likely that any mechanism set which includes off-LMSR operations (e.g., the margin call mechanic in~\ref{ssec: EL as short selling} and the phantom purchase in~\ref{ssec: EL as phantom shares}) would have a good chance of breaking the arbitrage-free nature of ordinary LMSR. 

It is possible that something like this may still work; we have not comprehensively explored the space.
However, the design space is complicated.
Some things which must be considered:
\begin{itemize}
    \item How large of a phantom purchase is necessary to achieve the desired result?
    \item How are the phantom shares managed?
    That is:
    \begin{itemize}
        \item I.e., when Alice purchases $r$, some phantom shares are instantiated which seem to be somehow tied to those shares of $r$.
        What happens if Alice sells or transfers the associated shares of $r$?
        \item A similar issue exists with the original shares of $A$; is Alice allowed to sell or transfer her original shares of $A$?
    \end{itemize}
    \item If a Doubt/Mindchange mechanic is integrated here, how does that work?
\end{itemize}

\section{Research Agenda}
\label{sec: research agenda}
This section describes several avenues which should be pursued in the next research effort for Carroll Mechanisms.
We begin very broadly in Section~\ref{ssec: the top list} by describing the high-level goals which must be forefront; resolution on these goals will enable rigor on the remainder of the work.
Following this, we describe some research questions which focus more specifically on Epistemic Leverage and the Restake mechanism in Section~\ref{ssec: questions}.
We close with Modeling Notes (Section~\ref{ssec: modeling notes}) and Mechanism Notes (Section~\ref{ssec: mech notes}) to discuss some specific ideas which could be pursued to advance the agenda.

\subsection{The Immediate Overarching Agenda}
\label{ssec: the top list}

The Fall 2025 project has yielded some stunning success in some dimensions (e.g., the combinatorial LMSR formulation of networked prediction markets), but there are certain core ambiguities which are beginning to hinder forward progress.
In this section, we expand upon the list from Page~\pageref{list: 3 key pieces} which delineated three core components which must be the first line of inquiry of any future research.
Substantial progress on these pieces will greatly facilitate future research.

\subsubsection{A clear and concrete statement of system objectives}
\label{sssec: objectives}
This piece has been elusive since the beginning of the Fall 2025 project.
This needs to incorporate a careful process, conceptually linked in a sequence like this:
\begin{itemize}
    \item {\bf Narrative: frame the problem.}
    The failings of incumbent systems must be framed and specific failings must be singled out for improvement.
    For example: in token voting, whales make all the decisions simply by merit of the size of their wallets.
    Futarchy has the property that metrics and decision options must be defined up front; in addition, it's inherently susceptible to Othman-Sandholm manipulation~\cite{Othman2010}.
    Neither of these have any in-built affordance which allows traders to make statements about their reasons for taking a position.
    Which of these problems specifically is the proposed system attempting to mitigate?
    The original proposal and community discussions contain a lot of this material, but research time needs to be devoted to these questions to synthesize the various concepts into a cohesive narrative.
    This work will likely be an extended sequence of discussions between Connor and a researcher, and should result in a structured (e.g., tabular) list of incumbent systems with clear descriptions of their failings. 
    
    \item {\bf Qualitative: articulate how a successor system should behave.}
    If the \textit{Narrative} segment explains where existing systems fail, this segment should explain (qualitatively) what a successor system \textit{succeeding} would look like.
    For example: the Fall 2025 project had a theme of attempting to incentivize a notion of \textit{sincerity} in prediction markets; this was a significant portion of the Restake/Doubt/Mindchange mechanics.
    However, we never formally specified which of the incumbent shortcomings are meant to be mitigated by this proposed property.
    Again, the output should be a structured natural-language list of desirable properties that are each very clearly tied to the Incumbent Shortcomings from the \textit{Narrative} segment.
    Care should be taken to make these property definitions as modular as possible; any unavoidable complementarities between properties should be clearly identified.
    
    \item {\bf Quantitative: how will be know we've succeeded?}
    What metrics (or formal properties) does the proposed system need to optimize (possess)?
    Once the \textit{Qualitative} segment is complete and a system's desirable properties have been sufficiently articulated, it will be necessary to specify formal success criteria.
    This too will cost substantial research time because it is generally hard to identify formal metrics and properties, even when the system goals already have clear natural-language descriptions.
    Where possible, existing metrics/properties from the literature should be applied (for example, \textit{truthfulness}) to promote coherence with existing work.
    In addition, the relevant metrics/properties will depend greatly on the selected meta-architecture.
    Thus, we expect that this will be tightly integrated or iterative with the meta-architecture and literature review (Sections~\ref{sssec: meta-architecture},~\ref{sssec: lit review}).
\end{itemize}
Practically speaking, the \textit{Narrative}, \textit{Qualitative}, and \textit{Quantitative} segments of this step will be informed by one another and may not be executed strictly sequentially.
Nonetheless, we believe that thinking about these as distinct and necessary pieces of the research program will promote clarity in the resulting statement of goals.

To seed the effort, here we present several properties which we discussed during the course of the Fall 2025 project.
In most cases, we have not successfully tied these informally-stated properties to specific objectives.
Some of these properties have been:
\begin{itemize}
    \item {\bf Relevance Elicitation:}
    The key idea here is that we want the mechanism set to have the intrinsic ability to \textit{reason} by logically connecting various propositions.
    However, the specific meaning of this property has never been totally clear.
    Does it mean that prices between linked propositions move together?
    Does it mean that traders honestly divulge what they believe about relevance?
    It's been a very slippery concept, in part due to our unclear understanding of how we might get resolution on a relevance prediction market.
    In addition, we need to tie concepts like this to higher-level meta-objectives.
    \item {\bf Long-Range Propagation:}
    We have defined this vaguely to mean something like ``price movements in one part of the market can propagate to other propositions in the market, potentially over a long distance.''
    We can loosely claim that our proposed mechanism set possesses this property simply due to the networked structure, but the property has never been posed rigorously or connected to any higher-level meta-objectives.
    That is, the ``narrative'' and ``qualitative'' segments for this property are totally missing.
    \item {\bf Degradation to Futarchy:}
    This is a clearer objective; ideally, we imagine having a tuning parameter which can smoothly dial our mechanism all the way down to Futarchy.
    Because we have never completely settled on a mechanism set, we have never checked specifically whether this is satisfied; however, it does seem the simplest of these properties to satisfy.
    One might think of this as a conditional property: it is only a relevant property if a decision market can be implemented in our system.
    Thus, requiring this property may imply that we are requiring some specific set of meta-architectures.
    \item {\bf Proper Scoring:}
    Because we have always built on top of combinatorial LMSR, we have taken this property for granted --- but as with all, we need to be careful.
    This needs to be stated more formally when we do not have resolvable markets.
    It will likely be instructive to see how the issue is handled by self-resolving markets~\cite{Srinivasan2025} and adopt certain aspects of that approach.
    In addition, the Proper Scoring property would almost certainly be lost if extra affordances are added on top of basic LMSR trades.
\end{itemize}

\subsubsection{A clear and concrete meta-architecture}
\label{sssec: meta-architecture}
In my informal notes, I call this the ``end-caps'' question: how does information get in to the system, and how do actions come out of the system?
For example, are the nodes in the Carroll structure \textit{an input to a decision market}, or are they themselves \textit{literally a decision market?}
Do all nodes need to be self-resolving, or do some come with a trusted resolution source?
Most likely, this will be tightly integrated with the Section~\ref{sssec: objectives} objectives work: it's difficult to know what objectives are relevant without knowing how the system interfaces with the world.

The approach in the Fall 2025 project was largely to leave this ambiguous for the sake of generality; however, the lack of defined end-caps hurt us in the end because it led to ambiguous assumptions regarding the trader incentive problem.
Accordingly, I recommend that some research/planning time be devoted to deciding on a specific meta-architecture so that incentive properties can be clearly framed.
Here, it matters more to commit to a particular architecture than it does which particular architecture is selected (of course, provided that the selected architecture is judged to be sufficiently relevant to the objectives under consideration from Section~\ref{sssec: objectives}).
It is possible that this commitment will result in the loss of some generality; I would argue that this is an acceptable tradeoff since it will facilitate a much more rigorous level of reasoning.
When two different architectures are under consideration, I recommend ties be broken in favor of simpler ones for the sake of complexity reduction.

In the Fall 2025 project, several potential things have surfaced relevant to this concept.
We deal with some of these in more detail elsewhere, but a summary is as follows:
\begin{itemize}
    \item ``Every node is a prediction market'': this was the implicit assumption underlying all of the combinatorial LMSR work.
    This approach {essentially} fails immediately due the lack of a resolution source for either a decision node (e.g., ``school should start at 9 am'') or a relevance edge (a paired proposition, in the language of Section~\ref{sec:foundation}).
    \item The top-level nodes are a decision market (in the sense of~\cite{Othman2010}, where the nodes are conditional prediction market nodes tied to events like ``metric X will be achieved given decision A,'' ``metric Y will be achieved given decision B''), and there is a logic layer on top of this which outputs a decision as a function of the market's predictions.
    \item There's a concept from Connor that a Carroll market is somehow an \textit{input} to a decision market of some kind; that the Carroll mechanism allows users to adjudicate what the metrics and decisions should be in some kind of Futarchic system.
\end{itemize}

The reason this step is crucial is that the incentives seen by traders matter; if the market renders decisions, traders' behavior will be impacted by the specific mechanism by which decisions are rendered.
In this sense, we might call this the ``meta-mechanism'' section: Carroll mechanisms ``live inside'' a bigger system which \textit{interacts with the world somehow}; we need a very clear concept of how these external interactions actually work so that it is clear what engineering/incentive challenges we're actually trying to solve.

\subsubsection{A more extensive literature review}
\label{sssec: lit review}
This requires very little explanation.
In Fall 2025, the bulk of the progress here was in a day or two when we discovered the power of combinatorial prediction markets, and then again during the creation of this document.
My practical recommendation is to start with~\cite{Srinivasan2025} (published in 2025, deals with a similar problem to some of ours, has a very extensive literature review on a broad range of subjects beyond classical prediction markets) and dig deeply from there.
It also may be useful to explore the epistemic game theory literature a bit more deeply to try to understand which modeling approaches are ``correct'' given the problem we are trying to solve, whether or not these approaches end up being practical or tractable.

Furthermore, the researcher should commit to revisiting this regularly throughout the course of the research project, perhaps spending a day every two or three weeks to go back to the literature to look for work relevant to any new concepts which have been uncovered.
This can help to avoid certain research failure modes in which the researcher suddenly thinks of a ``new'' angle but fails to realize that it has already been considered (either positively or negatively) by others.

\subsection{Key Questions for Epistemic Leverage}
\label{ssec: questions}
After establishing the main technical foundation of combinatorial LMSR, the main focus of the Fall 2025 project was the concept of Epistemic Leverage (EL).
We recommend that no further research work be devoted to EL until the Section~\ref{ssec: the top list} work is much further advanced.
However, once this is complete, here we outline a few key agenda points for studying EL in more depth.

\subsubsection{Goal: State Epistemic Leverage as a \textit{Property.}}
\label{question: possible}

It would be very nice to pose epistemic leverage as a \emph{property} of a mechanism.
In natural language, the property might be stated as something like
\begin{quote}
    Epistemic Leverage means that agents who \textit{honestly} declare the reasons for their votes have more influence than those who are (1) silent on their reasons, or (2) dishonest about their reasons.
\end{quote}
If such a property were posed in a rigorous, formal way, then in principle any mechanism could be checked to see if it possesses the property.
Alternatively, one could assume that an EL property holds and then derive attributes/properties of resulting mechanisms, leading to the possibility of deriving negative results.

\subsubsection{Question: \textit{why} do we want to participants to reveal their reasoning?}
In other words, we need to tie our desire for EL to a higher-level objective. 
We have essentially assumed during Fall 2025 that EL is desirable without having a clear explanation of \textit{why;} if we could develop such an explanation, it would add substantial clarity to the overarching objectives and add significantly to the problem formulation (Section~\ref{ssec: the top list}).

Somehow, we're looking for {\em intellectual honesty} in governance: can we ``tilt the scales'' toward empowering people who tell us why they're voting how they're voting?
\begin{itemize}
	\item What does it mean to be \textit{intellectually honest}?
	In real life, it is almost a requirement for someone to say why they are taking their position: Senator X has some spiel about why they voted Yes on Bill A.
	However, it can be very difficult to understand (a) whether they actually believe that, and (b) whether it is actually an argument that holds water.
	\item Thus, Epistemic Leverage \emph{isn't} about rewarding people who simply tell us why they're voting how they're voting.
	That's a really important thing to note: talk is cheap!
	\item Then here are the questions:
	\begin{itemize}
		\item \textit{If we had a resolution mechanism for relevance claims}, do there exist mechanisms which empower intellectual honesty? See discussion in Section~\ref{ssec: modeling notes}, ``Temporarily assume away the hard part.''
		\item Much harder: How do we build a resolution mechanism for relevance claims (perhaps self-resolving prediction markets~\ref{sssec: self-resolving})?
		\item Is there any way to solve these 2 problems independently? (I suspect not; it has to be life-cycle because the working of the resolution mechanism depends so finely on the specific goals of the participants.)
	\end{itemize}
\end{itemize}

\subsubsection{Key Question: Is Reputation Required?}
\label{question: reputation}
{\bf My Hunch: Almost Certainly.}\\
Is there a way to give additional power/signal to ``sincere'' traders without requiring their reputation to come into play?
This question may be fundamentally about the Doubt mechanic, which is an affordance which allows people to place bets on (vote on?) whether they believe in someone's reasoning.
This is tightly connected to the challenge of asymmetric sincerity verification, as in Section~\ref{challenge: asymmetric sincerity}.

My hunch that reputation is required is driven by the fact that all participants (both honest and dishonest) have access to the same set of tools, and it is very difficult to automatically identify certain kinds of dishonesty.
Most likely, there are certain kinds of dishonesty/insincerity that will be impossible for the mechanism to identify but relatively easy for participants to identify (e.g., restaking on a false but irrelevant proposition).
Something tied to reputation is a clear way to address this, but integrating it wisely may be difficult.
It seems that the desired functionality may look something like voting or betting on whether particular batches of stake are honest or not, but then this is simply another mechanism which could be exploited for selfish purposes.

\subsubsection{Key Low-Level Question: Does Restake do what it needs to do?}
\label{question: restake properties}
{\bf Intuition: Restake must satisfy several key properties:~\ref{req: sincere restake},~\ref{req: spurious restake},~\ref{req: double vindication},~\ref{req: irrelevant restake proof}.}\\
Essentially what these properties all say is that any interpretation of the $A/r/B$ Restake concept must only have a special positive effect on $A$ when $B$ falls and $r$ rises; if $B$ is nondecreasing or $r$ is nonincreasing, restake should not help $A$.
Failing any of these properties enables a trader to gain an unwarranted advantage by using Restake.
Any interpretation of Restake must be verified to satisfy these properties.

\subsubsection{Key Low-Level Question: Can Restake be Initalization-Attack-Proof?}
\label{question: initialization attack}
{\bf Intuition: the RIA property (\ref{req: restake initialization proof}) provides a family of upper bounds on the restake bonus.}\\
The Restake Initialization Attack is a dangerous attack on the restake mechanism which attempts to prime the market for conditions which would provide an unwarranted restake bonus.
The attack vector is available at all times (i.e., the attacker doesn't have to wait for nice conditions) and exploits the \textit{core} use-case for Restake.

Any interpretation of restake has to address this attack. 
The RIA property~(\ref{req: restake initialization proof}) explicitly forces this attack to be unprofitable, but the exact consequences of satisfying RIA will depend on the particular realization of Restake under consideration.
It is anticipated that satisfying this property will imply a family of fundamental upper bounds on the restake bonus (or a family of fundamental lower bounds on the cost of a restake contract).

\vspace{2mm}
\begin{mdframed}\vspace{-2mm}
\subsubsection*{Preliminary Exploration of RIA Property}

This property is critical enough that I will sketch out a specific agenda. 
The attacker model is tricky.
The attacker has $c$ capital, and is willing to lose $\ell\leq c$ capital to achieve its goal of boosting $A$.
The baseline case is that the attacker simply invests all $c$ in proposition $A$, resulting in the honest $A$-signal $p_A$.
The alternative case is that the attacker splits their buy into a portion $c_A$ for a restake-enabled investment in $A$, and a portion $c_B$ for a temporary boost of $B$, where $c_A+c_B=c$.
This $c_B$ is invested directly into $B$, then the $c_A$ is invested in an $A/r/B$ restake, then the $B$ position is immediately exited.
The proceeds (if any) of the sale of B shares can then be invested in $A$ as well, for a final $A$-signal of $p_A'$.
It seems that the RIA property says that for all choices of $(c_A,c_B)$, it must be the case that
$$p_A\geq p_A'.$$

\noindent {\bf Code results for phantom-share-restake:} 
See the {\tt ria.py} script in the code delivery. 
The {\tt frac\_to\_leave} and {\tt A\_frac} variables can be tuned to adjust how much capital the attacker saves for restake, and splits between $A$-purchase and $B$-manipulation; the final 2 market prints correspond to no-attack and attack, respectively.
Preliminary experiments appear to indicate that if the restake model is akin to the phantom share model in~\ref{ssec: EL as phantom shares}, then RIA-proofness is satisfied under this definition.

\end{mdframed}

\subsection{Modeling Notes}
\label{ssec: modeling notes}
In this section we collect miscellaneous notes on possible ways to model the problem.

\subsubsection{Temporarily assume away the hard part.}
    This would significantly narrow the scope and could result in some negative results.
    The fact is that we do not currently know how to do relevance elicitation. 
    But suppose we did --- in that case would Carroll Mechanisms give us what we want? 
    I.e., one strategic environment that might be interesting initially is one in which we assume that relevance is resolvable, and then we use that as a first-order check of whether the mechanism set is useful. 
    If the mechanism set isn't useful given resolvable edges, then there's no point digging deeper to figure out how to resolve edges.
\subsubsection{Is Doubt totally and simply connected to reputation?}
    Somehow Doubt is an action which says ``I don't think you'll change your mind,'' or ``I think you're being insincere.''
    These might be clearly connected to reputation.
    So: if doubt is sort of a deterrent, then maybe only sincere people restake?
    In this case, do we get good outcomes in the sense that \emph{because} restaking becomes this extra tool which gives sincere people more power, they're now able to fight the man.

\subsubsection{Is there a place for heterogeneous priors?}
It has always been an open question of how to handle trader beliefs.
In the few times that I have begun to write a specific formal model, it has seemed reasonable to think about traders as having a prior belief over the set of possible truth values of all propositions.
But then how do we deal with traders that fundamentally disagree about the way the world works?
There must be some framework from epistemic game theory which handles this; probably we need to incorporate this into the plan for literature review.

A related modeling note has to do with ``mindchange'': in a Bayesian common prior world, the phrase ``changing your mind'' is almost a misnomer.
If a Bayesian agent experiences a large change in belief from one posterior to the next, their new belief must still be consistent with something they believed was possible \textit{a priori}.
I.e., they haven't so much ``changed their mind'' as much as re-allocated their existing beliefs. 
Perhaps this might result from seeing a signal which implies a particularly low-probability state in your prior, but there is no sense in which the new information somehow now makes you think about the world differently.
You have simply received information that amplifies something you knew and believed possible all along.

But when we talk about changing your mind, it really feels very much like we mean something like \textit{adopting a new prior}, which a Bayesian agent cannot do.
Again, I think it would be worthwhile to spend some time in the epistemic game theory literature to find some kind of reasonable way to think about this.
Asking this question may open an intractable can of worms; however, it would at least be useful to have some clarity on what modeling approaches might be relevant.

\subsection{Mechanism Notes}
\label{ssec: mech notes}
In this section we collect miscellaneous notes on hypothetical mechanism components to solve the problem.

\subsubsection{Self-Resolving Prediction Markets:}
\label{sssec: self-resolving}
The EC'25 paper which proposed self-resolving prediction markets~\cite{Srinivasan2025} is the clearest picture I've yet seen of something which has a chance at solving the relevance elicitation problem.
The mechanism does rely on some assumptions which worry me:
\begin{itemize}
    \item Traders are Bayesian with a common prior over \emph{all} signals (and as you'd expect from that assumption, are highly computationally capable),
    \item Traders are outcome-agnostic,
    \item Predictions do not impact outcomes (non-performativity),
    \item Every agent participates exactly once (although this \textit{might} be without loss of generality),
    \item We have a source of randomness.
\end{itemize}
The paper relies on the famous Aumannian concept~\cite{Aumann1976} that ``agents can't agree to disagree,''  (1-sentence primer: if 2 agents have the same prior, then if they \textit{know} each others' posterior [i.e., they understand each others' experience of the world], they must \textit{have} the same posterior [i.e., agree with each other about the state of the world].).
Assuming sincerity and a common prior feels like a risky framework in a context where we're particularly worried about possibly-insincere traders who have heterogeneous priors.

Nonetheless, it would be an interesting exercise to think through how self-resolving markets might be incorporated to do relevance elicitation (i.e., to resolve paired propositions in the Carroll structure).
The self-resolving mechanism works by using the last trader as the global reference for all previous traders; clearly, this trader is extremely powerful.
The clever trick to prevent this trader from manipulating the market is that the market closes to new trades at a random time --- thus, the reference trader cannot know they're the reference trader until it is too late to exploit their advantage.

Essentially, you're so unlikely to be the reference trader that you can effectively assume that you're \textit{not} the reference trader.
Thus, if self-resolving markets were used for relevance elicitation, they would necessarily resolve at an unpredictable time, and then presumably the relevance values would be fixed for the remainder of the interaction.%
\footnote{Although the paper does note (p. 562) that you might be able to get away with a rolling window where new reference agents are selected (randomly) every $T$ agents \textit{in expectation.}
This has the downside of requiring unbounded payouts.}

I think one of the tricky parts of the self-resolving concept would be to do it in a permissionless way. 
There is this critical parameter $k$, the \textit{number of informational substitutes that the reference agent knows that none of the non-reference agents can access.}
The concept in the paper is that every agent knows something that other agents don't; we get some form of incentive compatibility as long as the reference agent knows $k$ \textit{more} things than any of the other incentivized agents.
To make sure this happens, we simply don't incentivize any of the last $k$ agents (instead, pay them a flat fee); this ensures that the reference agent (last agent) knows $k$ more things than any of the others (because the reference agent learned each of the things that the last $k$ agents revealed).

In our case, if we don't have identities, all we can do is say something about the last $k$ \textit{trades} rather than the last $k$ traders.
This may dramatically enhance adversaries' ability to manipulate the mechanism, because you would lose the notion of conditional independence on the signals that are being reported in subsequent trades.

Another issue is that of heterogeneous priors (which feels like the right way to model disagreement on fundamentals).
There's an interesting statement (the bottom of p. 555) that a reference agent (or I suppose any subsequent trader) can check if a previous agent's report is consistent with their own information structure, and just ignore it if not.
Then imagine a scenario where there are two populations of traders; each population has its own internal ``common'' prior.
If each population doesn't know the other's prior, what happens?
It might be plausible for an agent from population 1 to simply write off the traders from population 2, chalking them up to idiots or noise traders; i.e., naively, you might expect the population to simply bifurcate into two groups which each push the market back and forth consistent with their own prior.
However, this probably isn't right unless all traders are really ignorant about the possibility of someone else's prior: the argument for incentive compatibility in~\cite{Srinivasan2025} does actually depend on the reference agent behaving properly.
Suppose you picked one specific reference agent and this agent were from Population 1: then if it were true that the population bifurcates, this agent would simply ignore the reports from Population 2 and go along with Population 1's prior.
But then all the Population 2 traders would be scored and paid according to the Population 1 prior, and they wouldn't be happy with this result, meaning that it cannot be an equilibrium for one population to simply ignore the other.
I \textit{think} this simply means that the mechanism is not generally truthful under heterogeneous priors, which feels like a problem.
But maybe that is not too big a problem; even with wildly diverging priors you might just get an equivalence with token voting, or perhaps it would be meaningful to incorporate some mechanic that only resolves if the variance of the last $k$ prices is small, or something like that.
Another potentially interesting research question is something like ``how different do the priors need to be for the mechanism to cease to be truthful.''
One might expect that 2 populations of traders with very similar priors might still retain a degree of truthfulness.

Finally, there's a brief note in the paper (p. 562) that if the mechanism designer has access to trusted agents, these agents can be placed at the end and made into the reference agents.
This seems difficult to implement, but perhaps could integrate somehow with the notion of being ``feared'' for being intellectually honest.

\subsubsection{Restake is a pre-commitment to switch $A\to B$ at a predefined price:} Instead of making Mindchange a thing that has to be triggered, what if it were just automatic once you get to a particular price?
    This would precisely align with the concept of ``I would change my mind if....''
    It would probably go well with a locking mechanism on Restake; you cannot exit a restake position, but you can toggle between A and B within the position.
    It slightly re-interprets restake as a statement that ``I think these 2 are linked; I prefer A, but I'll go where the evidence takes me.''
    It sounds spiritually similar to what we want!

\subsubsection{Restake is a contract that buys $A$ later:} 
When a trader restakes, this places cash into a contract; this contract then executes (purchasing $A$ at a predefined price) when $B$ falls below a threshold.

\subsubsection{Restake is ``buy shares in $A\land r$'':} What is missing if we implement Restake simply as ``buy shares in $A\land r$?'' 
    Buying $A\land r$ is a risky move for anyone who does not actually believe $r$ (or who actually believes $B={\tt True}$).
    It gives a huge bang for the buck if executed when $B$ is high and $r$ is low and then they swap, which automatically satisfies the Double Vindication property of restake (Section~\ref{req: double vindication}).

\section{Conclusion}
\label{sec:conclusion}

The project has yielded some substantial understanding of the problem space in general, and also of potential approaches to the problem.
In addition, we have identified significant areas of challenge which will need to be surmounted.
Finally, we have clearly framed a research agenda to set the stage for a productive next phase of the research.











%
\bibliographystyle{IEEETran}
\bibliography{PhilipMendeley}

@article{Aumann1976,
    title = {{Agreeing to Disagree}},
    year = {1976},
    journal = {The Annals of Statistics},
    author = {Aumann, Robert J.},
    number = {6},
    month = {11},
    pages = {1236--1239},
    volume = {4},
    url = {https://projecteuclid.org/journals/annals-of-statistics/volume-4/issue-6/Agreeing-to-Disagree/10.1214/aos/1176343654.full},
    doi = {10.1214/aos/1176343654},
    issn = {0090-5364}
}

@article{Hanson2003,
    title = {{Combinatorial Information Market Design}},
    year = {2003},
    journal = {Information Systems Frontiers},
    author = {Hanson, Robin D},
    number = {1},
    month = {1},
    pages = {107--119},
    volume = {5},
    publisher = {Kluwer Academic Publishers},
    url = {https://link.springer.com/10.1023/A:1022058209073},
    doi = {10.1023/A:1022058209073},
    issn = {1387-3326}
}

@inproceedings{Chen2008a,
    title = {{Complexity of combinatorial market makers}},
    year = {2008},
    booktitle = {Proceedings of the 9th ACM conference on Electronic commerce},
    author = {Chen, Yiling and Fortnow, Lance and Lambert, Nicolas and Pennock, David M. and Wortman, Jennifer},
    month = {7},
    pages = {190--199},
    publisher = {ACM},
    url = {https://dl.acm.org/doi/10.1145/1386790.1386822},
    address = {New York, NY, USA},
    isbn = {9781605581699},
    doi = {10.1145/1386790.1386822}
}

@article{Olfati-Saber2007,
    title = {{Consensus and cooperation in networked multi-agent systems}},
    year = {2007},
    journal = {Proceedings of the IEEE},
    author = {Olfati-Saber, Reza and Fax, J. Alex and Murray, Richard M.},
    number = {1},
    pages = {215--233},
    volume = {95},
    isbn = {0018-9219},
    doi = {10.1109/JPROC.2006.887293},
    issn = {00189219},
    pmid = {4118472},
    arxivId = {1009.6050}
}

@article{Saber2003,
    title = {{Consensus protocols for networks of dynamic agents}},
    year = {2003},
    journal = {Proceedings of the 2003 American Control Conference, 2003.},
    author = {Saber, R.O. and Murray, Richard M.},
    pages = {951--956},
    volume = {2},
    url = {http://ieeexplore.ieee.org/document/1239709/},
    isbn = {0-7803-7896-2},
    doi = {10.1109/ACC.2003.1239709},
    issn = {0743-1619}
}

@inproceedings{Chen2011,
    title = {{Decision Markets with Good Incentives}},
    year = {2011},
    booktitle = {International Workshop on Internet and Network Economics (WINE)},
    author = {Chen, Yiling and Kash, Ian and Ruberry, Mike and Shnayder, Victor},
    pages = {72--83},
    url = {http://link.springer.com/10.1007/978-3-642-25510-6_7},
    doi = {10.1007/978-3-642-25510-6{\_}7}
}

@inproceedings{Othman2010,
    title = {{Decision Rules and Decision Markets}},
    year = {2010},
    booktitle = {AAMAS '10: Proceedings of the 9th International Conference on Autonomous Agents and Multiagent Systems},
    author = {Othman, Abraham and Sandholm, Tuomas},
    month = {5},
    pages = {625--632},
    url = {www.ifaamas.org},
    isbn = {9780982657119},
    doi = {10.5555/1838206.1838288}
}

@incollection{Hossain2025,
    title = {{Designing Automated Market Makers for Combinatorial Securities: A Geometric Viewpoint}},
    year = {2025},
    booktitle = {Proceedings of the 2025 Annual ACM-SIAM Symposium on Discrete Algorithms (SODA)},
    author = {Hossain, Prommy Sultana and Wang, Xintong and Yu, Fang-Yi},
    month = {1},
    pages = {1329--1365},
    publisher = {Society for Industrial and Applied Mathematics},
    url = {https://epubs.siam.org/doi/10.1137/1.9781611978322.40},
    address = {Philadelphia, PA},
    doi = {10.1137/1.9781611978322.40}
}

@article{Chen2010,
    title = {{Designing Markets for Prediction}},
    year = {2010},
    journal = {AI Magazine},
    author = {Chen, Yiling and Pennock, David M},
    number = {4},
    month = {12},
    pages = {42--52},
    volume = {31},
    url = {https://onlinelibrary.wiley.com/doi/10.1609/aimag.v31i4.2313},
    doi = {10.1609/aimag.v31i4.2313},
    issn = {0738-4602}
}

@article{Chen2014a,
    title = {{Eliciting Predictions and Recommendations for Decision Making}},
    year = {2014},
    journal = {ACM Transactions on Economics and Computation},
    author = {Chen, Yiling and Kash, Ian A. and Ruberry, Michael and Shnayder, Victor},
    number = {2},
    month = {6},
    pages = {1--27},
    volume = {2},
    url = {https://dl.acm.org/doi/10.1145/2556271},
    doi = {10.1145/2556271},
    issn = {2167-8375}
}

@inproceedings{Oesterheld2023,
    title = {{Incentivizing honest performative predictions with proper scoring rules}},
    year = {2023},
    booktitle = {39th Conference on Uncertainty in Aritifical Intelligence (UAI 2023)},
    author = {Oesterheld, Caspar and Treutlein, Johannes and Cooper, Emery and Hudson, Rubi},
    number = {216},
    pages = {1564--1574},
    volume = {PMLR},
    issn = {26403498}
}

@article{Hanson2007,
    title = {{Logarithmic Market Scoring Rules for Modular Combinatorial Information Aggregation}},
    year = {2007},
    journal = {The Journal of Prediction Markets},
    author = {Hanson, Robin D},
    number = {1},
    month = {12},
    pages = {3--15},
    volume = {1},
    url = {http://www.bjll.org/index.php/jpm/article/view/417},
    isbn = {7039932326},
    doi = {10.5750/jpm.v1i1.417},
    issn = {1750-6751}
}

@inproceedings{Pennock2011,
    title = {{Price updating in combinatorial prediction markets with Bayesian networks}},
    year = {2011},
    booktitle = {Proceedings of the 27th Conference on Uncertainty in Artificial Intelligence, UAI 2011},
    author = {Pennock, David M and Xia, Lirong},
    pages = {581--588},
    arxivId = {1202.3756}
}

@inproceedings{Srinivasan2025,
    title = {{Self-Resolving Prediction Markets for Unverifiable Outcomes}},
    year = {2025},
    booktitle = {Proceedings of the 26th ACM Conference on Economics and Computation},
    author = {Srinivasan, Siddarth and Karger, Ezra and Chen, Yiling},
    month = {7},
    pages = {547--573},
    publisher = {ACM},
    url = {https://dl.acm.org/doi/10.1145/3736252.3742593},
    address = {New York, NY, USA},
    isbn = {9798400719431},
    doi = {10.1145/3736252.3742593}
}

@inproceedings{Chakraborty2016,
    title = {{Trading on a rigged game: outcome manipulation in prediction markets}},
    year = {2016},
    booktitle = {IJCAI'16: Proceedings of the Twenty-Fifth International Joint Conference on Artificial Intelligence},
    author = {Chakraborty, Mithun and Das, Sanmay},
    pages = {158--164}
}

\end{document}